# Phase lags and coherence of X-ray variability in black hole candidates

Michael A. Nowak[1] and Brian A. Vaughan[2]

[1]*JILA, Campus Box 440, Boulder, CO  80309-0440*
[2]*Caltech, Pasadena, CA  91125*

## ABSTRACT

The "low" (hard or "non-thermal") state of black hole candidates is sometimes modelled via an optically thick, hot Compton cloud that obscures a softer input source such as an accretion disk. In these models the observed output spectra consist entirely of photons reprocessed by the cloud, making it difficult to extract information about the input spectra. Recently Miller (1995) has argued that the Fourier phase (or time) lag between hard and soft X-ray photons in actuality represents the phase lags intrinsic to the input source, modulo a multiplicative factor. The phase lags thus would be a probe of the input photon source. In this paper we examine this claim and find that, although true for the limited parameter space considered by Miller, the intrinsic phase lag disappears whenever the output photon energy is much greater than the input photon energy. The remaining time lags represent a "shelf" due to differences between mean diffusion times across the cloud. As pointed out by Miller, the amplitude of this shelf – which is present even when the intrinsic time lags remain – is indicative of the size and temperature of the Compton cloud and is a function of the two energies being compared. However, we find that with previous instruments such as Ginga the shelf, if present, was likely obscured by counting noise. A more sensitive measure of Compton cloud parameters may be obtainable from the coherence function, which is derived from the amplitude of the Fourier cross power spectral density. This function has been seen to exponentially decrease at high Fourier frequencies in Cygnus X-1. Coherence loss is characteristic of Compton clouds that undergo large variations of size and/or temperature on time scales longer than about 10 seconds. We argue that observing phase lags and coherence at a variety of energies and Fourier frequencies is more likely to reveal information about the nature of Compton clouds, rather than about their soft input sources.

## 1 INTRODUCTION

A number of black hole candidates (BHC) exhibit energy spectra that are well modelled by power laws with photon indexes of $\lesssim -1.7$ between energies of $\sim 1- \gtrsim 100$ keV (so-called "low states"). These spectra have been successfully modelled as soft input photons upscattered in an optically thick Compton cloud. The power law spectra is obtained from the fact that although the probability of scattering decreases exponentially with the number of scatters, the energy gain by the photons increases with this number. The two effects – with some tuning of parameters – balance each other, yielding the characteristic photon index of $\sim -2$ (*cf.* Sunyaev & Titarchuk 1980, Pozdnyakov et al. 1983, Titarchuk 1994). Two problems exist with this model, however. First, it is difficult to obtain information about the soft input spectrum from the hard output spectrum. Second, it is difficult to distinguish the predictions of the Compton cloud models from those of optically thin models (*cf.* Melia & Misra 1993, Luo & Liang 1994, etc.). One possible way to garner information about the input spectra and distinguish the Compton cloud models from optically thin models is to consider the copious variability data.

Black hole candidates in their low state tend to exhibit variability that is characterized by a Fourier power spectral density (PSD) that is inversely proportional to Fourier frequency, $f$. These sources with PSD $\propto f^{-1}$ also tend to show Fourier phase differences between hard and soft X-rays that are consistent with the hard photons lagging behind the soft photons (*cf.* Miyamoto et al. 1988, Miyamoto et al. 1991). The hard time lag, which is obtained by dividing the Fourier phase difference by $2\pi f$, increases nearly linearly with increasing Fourier period (Miyamoto et al. 1988). Miyamoto et al. (1988) argued that this casts doubt on the simple optically thick cloud model, for which one naively expects a constant time lag on the order of the diffusion time through the cloud. Recently, Miller (1995) has argued that the cloud does *not* greatly affect the intrinsic phase lags of the soft input source. He reduced the Comptonization problem to that of determining the "redistribution function", $R(E, E')\, dE$, which is the probability of scattering a photon of input energy $E'$ to output energy $E$. By ignoring the small phase induced by the mean diffusion time through the cloud and by assuming that the intrinsic phase differences between input channels was small, Miller found that the output phase lags were equal to the input phase lags – modulo a mulitplicative factor of order unity.

In this paper we examine this claim. Our approach is similar to that of Miller; however, we take more careful consideration of how the observations are actually made. Miller (1995) considered continuous transforms of continuous functions (specific intensities). Here we consider discrete transforms of discrete functions (observed photon counts). Miller also considered phase differences between discrete energies. Here we consider phase differences between energy bands with widths typical of X-ray observations. In addition, we consider the effect of scattering on the PSD. In general, scattering decreases the amplitude of the PSD. This was mentioned by Miller (1995), and it was studied in other works (*cf.* Kylafis & Klimis 1987, Brainerd et al. 1987, Bussard et al. 1988, Miller & Lamb 1992). Our approach is not as detailed as a number of these works, but it captures most of the qualitative and quantitative behavior. We can afford this simpler approach since, as Miller (1995) correctly pointed out, observing a given output energy band essentially fixes the mean diffusion time through the cloud. The degradation of the PSD then arises from the variance in the diffusion time *for a specific set of input and output energies*. Finally, Miller (1995) was mainly concerned with stationary scattering clouds. We consider the possibility that the clouds have statistically fluctuating properties. This is seen to lead to a decrease in the coherence function at high Fourier frequencies. The coherence function is derived from the amplitude of the cross power spectral density between hard and soft photons, and it is a measure of the correlation between hard and soft photon variability (Vaughan 1991). Coherence loss at high frequency has been observed in Ginga data of the low state of Cygnus X-1, and may indicate that large amplitude variations do exist in Compton clouds (Vaughan 1991).

The outline of this paper is as follows. In §2 we calculate the effects of a stationary cloud on the input variability. In §3 we present a simple Compton cloud model that we use to develop a qualitative understanding of the mean photon diffusion time and its variance as a function of input and output photon energies. We then use this model in §4 to calculate output PSDs and time lags for blackbody input spectra. Although we only use a simple analytic Comptonization model, we expect the results to be qualitatively the same for more accurate numerical models. In §5 we present preliminary calculations of the effects of clouds with fluctuating parameters. We then summarize our results in §6.

## 2 EFFECT OF SCATTERING ON THE PSD & CPD

Let us assume we have a set of unscattered time series $g^i_{j[k]}$ that represent photon counts at times $t'_j$ and cover energy bands $[k]$. These time series are then scattered into a set of time series $h^i_{n[l]}$ that are measured at times $t_n$ and cover energy bands $[l]$. Mathematically, we can write this as

$$h^i_{n[l]} = F^i_{nm}[l] \ P^i_{mj}[kl] \ R^i_{[kl]} \ g^i_{j[k]} \ . \tag{1}$$

In the above, we use the convention of summation on repeated *subscripts* only. In addition, we have defined $R^i_{[kl]}$ as the matrix representation of the redistribution function (Miller 1995). Specifically, it is the fraction of photons scattered from energy band $k$ into energy band $l$. As Compton scattering preserves photon number, it is obvious that

$$\sum_l R^i_{[kl]} = 1 \ . \tag{2}$$

As we increase the number of energy bins $n$ to infinity, we regain Miller's definition of the redistribution function:

$$\lim_{n \to \infty} \langle R_{[kl]} \rangle = R(E, E') \ dE \ , \tag{3}$$

In equation (1), we also have defined $P^i_{mj}[kl]$ to be the fraction of photons scattered from energy band $[k]$ to energy band $[l]$ that is also scattered from time bin $j$ to time bin $m$. Finally, $F^i_{nm}[l]$ is the fraction of photons in energy band $[l]$ and time bin $m$ that is actually observed by the detector. We shall take this to be a diagonal matrix, and furthermore, we shall assume that the effects of counting noise are folded into this matrix. (Note that we are discounting the possibility of photons in a given energy band being misidentified by the detector as belonging to a different energy band.)

We have attached a superscript $i$ to our various matrices. This is because $F^i$, $P^i$, and $R^i$ are not true probability distributions, but rather a measure of what *actually* occured. However, as each time bin of the functions $g^i$ represents a large number of photons (the flux at the source), we expect $P^i$ and $R^i$ to be well defined by their underlying probability distributions for *each* input time series. On the other hand, since $F^i$ represents both Poisson noise and the reduction of the source flux to the level of the detected flux [typically as low as $\mathcal{O}(1)$ photon per time bin], it can be defined only by statistically averaging over a large number of time series. Furthermore, $g^i$ and $h^i$ themselves are assumed to be samples of statistically stationary processes. Therefore, their Fourier transforms are only statistically well defined by averaging over a large number of samples (*cf.* Davenport & Root 1987). The product of $F^i_{nm}[l]$ and $P^i_{mj}[lk]$ can be rewritten as

$$F^i_{nm}[l] \ P^i_{mj}[kl] = F[l] \ P^i_{nj}[kl] + N^i_{nj}[kl] \ , \tag{4}$$

where $F[l]$ represents the (assumed constant) response of the detector, and $N^i_{nj}$ represents noise. There exist a number of techniques for removing Poisson noise effects (or at least estimating their magnitudes) from power and cross power spectral densities (*cf.* Leahy et al. 1983, Vaughan 1991). We shall assume that such procedures will be carried out on the data sets of interest to us, and therefore from now on we will ignore the effects of noise.

We will assume that the probability distribution of the time delay induced by scattering is well represented by a Gaussian. That is,

$$P^i_{nj}[kl] \;=\; (2\pi\sigma_{kl}{}^2)^{-1/2} \; \exp\left(-\frac{[\,(t_n - t'_j) \;-\; \delta t_{kl}]^2}{2\sigma_{kl}{}^2}\right) \; \Delta \;, \tag{5}$$

where $\delta t_{kl}$ is the mean time delay induced by scattering from energy band $k$ to energy band $l$, $\sigma_{kl}$ is the standard deviation in this time, and $\Delta$ is the width of the (assumed equally spaced) time bins in both the input and output time series. We also shall find it convenient to define the quantity

$$\tau_{kl} \;\equiv\; 2\pi\sigma_{kl} \;. \tag{6}$$

We now introduce the discrete Fourier transform matrix $T$ and the inverse transform matrix $T^{-1'}$, defined by:

$$T_{pn} \;\equiv\; \exp\left(i\,2\pi\,f_p t_n\right) \; \Delta \;, \tag{7a}$$

$$T^{-1'}_{jo} \;\equiv\; \exp\left(-i\,2\pi\,f_o t'_j\right) \; \Delta^{-1} \;, \tag{7b}$$

(*cf.* Press et al. 1992). Defining $\mathcal{H}^i$ to be the transform of $h^i$ and $\mathcal{G}^i$ to be the transform of $g^i$, we can rewrite equation (1) as:

$$\mathcal{H}^i_{p[l]} \;=\; F[l]\, T_{pn}\, P^i_{nj}[kl]\, R_{[kl]} \left(T^{-1'}_{jo} \mathcal{G}^i_{o[k]}\right) \;. \tag{8}$$

Note that $T_{pn} P^i_{nj}[kl]$ yields the discrete Fourier transforms of the columns of $P^i$, and that these columns represent the probability distribution function of equation (5). Thus we have

$$\begin{aligned} T_{pn} P^i_{nj}[kl] \;&=\; \exp\left(i\,2\pi\,f_p t'_j\right)\, \exp\left(i\,2\pi\,f_p \delta t_{kl}\right)\, \exp\left[-(f_p \tau_{kl})^2/2\right]\, \Delta \\ &\equiv\; \mathcal{P}^i[p, kl]\, \exp\left(i\,2\pi\,f_p t'_j\right)\, \Delta \;=\; \mathcal{P}^i[p, kl]\, T_{pj}' \;. \end{aligned} \tag{9}$$

We have retained the superscript $i$ above to denote the fact that the probability distribution *does not* have to remain identical from sample to sample. (Though the above formulation *does* assume it remains stationary from column to column in $P^i_{nj}$. In principle, that assumption can be relaxed; however, the final forms of the equations will be more complicated.) Utilizing the identity $T_{pj}' T^{-1'}_{jo} = \delta_{po}$ (where $\delta_{po}$ is the Kronecker delta – valid for equally spaced time bins) equation (8) becomes

$$\mathcal{H}^i_{p[l]} \;=\; F[l]\, \mathcal{P}^i[p, kl]\, R^i_{[kl]}\, \mathcal{G}^i_{p[k]} \;, \tag{10}$$

or upon averaging over time series,

$$\langle \mathcal{H}_{p[l]} \rangle \;=\; F[l]\, \langle \mathcal{P}[p, kl]\, R_{[kl]} \rangle\, \langle \mathcal{G}_{p[k]} \rangle \;. \tag{11}$$

We have assumed that the input time series are statistically independent from the effects of the scattering cloud; however, we cannot in general assume that $\mathcal{P}[p, kl]$ and $R_{[kl]}$ are statistically independent.

For convenience, we shall combine $F[l]$, $P^i[p, kl]$ and $R^i_{[kl]}$ into a matrix which we shall call $\mathcal{Q}^i_{[kl]}(p)$. We then have for the mean power spectral density of $h_{j[m]}$

$$\langle |\mathcal{H}_{p[m]}|^2 \rangle \;=\; \sum_k \langle |\mathcal{Q}_{[km]}(p)|^2 \rangle\, \langle |\mathcal{G}_{p[k]}|^2 \rangle \;+\; \sum_{k \ne l} \langle \mathcal{Q}_{[km]}(p)\, \mathcal{Q}^*_{[lm]}(p) \rangle\, \langle \mathcal{G}_{p[k]} \mathcal{G}_{p[l]} \rangle \;, \tag{12}$$

and for the mean cross power spectral density (CPD) between $h_{j[m]}$ and $h_{j[n]}$

$$\langle \mathcal{H}_{p[m]} \mathcal{H}^*_{p[n]} \rangle = \sum_k \langle \mathcal{Q}_{km}(p) \mathcal{Q}^*_{kn}(p) \rangle \langle |\mathcal{G}_{p[k]}|^2 \rangle + \sum_{k \neq l} \langle \mathcal{Q}_{[km]}(p) \mathcal{Q}^*_{[ln]}(p) \rangle \langle \mathcal{G}_{p[k]} \mathcal{G}^*_{p[l]} \rangle . \quad (13)$$

In order to simplify notation, we shall drop the subscript $p$ and assume that all of our equations henceforth are implicitly functions of frequency. Furthermore, we shall drop the use of the averaging brackets and assume that all of our equations are averaged in accordance with equations (12) and (13).

The intrinsic transforms $\mathcal{G}_{[k]}$ can be written in terms of a real amplitude and real phase as

$$\mathcal{G}_{[k]} = \mathcal{A}_k \exp(i\phi_k) . \quad (14)$$

From this we can also define

$$\Delta \phi_{kl} \equiv \phi_k - \phi_l . \quad (15)$$

We shall also define the intrinsic coherence function

$$\mathcal{C}_{kl} \equiv \frac{|\langle \mathcal{G}_{[k]} \mathcal{G}^*_{[l]} \rangle|^2}{\langle |\mathcal{G}_{[k]}|^2 \rangle \langle |\mathcal{G}_{[l]}|^2 \rangle} . \quad (16)$$

The coherence function is $\leq 1$, and is a measure of how well correlated two energy bands are (Vaughan 1991). In order to further simplify the equations we shall absorb the factors of $F[l]$ into the normalizations of $\mathcal{H}_{[l]}$, and we shall define

$$\Delta \theta^{km}_{ln} \equiv 2\pi f \left( \delta t_{km} - \delta t_{ln} \right) . \quad (17)$$

Assuming a static scattering cloud (i.e. $\mathcal{P}^i$ does not change from time series to time series), we then have

$$|\mathcal{H}_{[m]}|^2 = \sum_k \exp\left\{-(f\tau_{km})^2\right\} R^2_{[km]} |\mathcal{G}_{[k]}|^2 +$$
$$\sum_{k \neq l} \exp\left\{-\frac{(f\tau_{km})^2 + (f\tau_{lm})^2}{2}\right\} \cos\left(\Delta\theta^{km}_{lm} + \Delta\phi_{kl}\right) R_{[km]} R_{[lm]} \sqrt{\mathcal{C}_{kl}} |\mathcal{G}_{[k]}||\mathcal{G}_{[l]}| ,$$
$$\mathcal{H}_{[m]} \mathcal{H}^*_{[n]} = \sum_k \exp\left\{-\frac{(f\tau_{km})^2 + (f\tau_{kn})^2}{2}\right\} \exp\left(i \Delta\theta^{km}_{kn}\right) R_{[km]} R_{[kn]} |\mathcal{G}_{[k]}|^2 +$$
$$\sum_{k \neq l} \exp\left\{-\frac{(f\tau_{km})^2 + (f\tau_{ln})^2}{2}\right\} \exp\left[i \left(\Delta\theta^{km}_{ln} + \Delta\phi_{kl}\right)\right]$$
$$\times R_{[km]} R_{[ln]} \sqrt{\mathcal{C}_{kl}} |\mathcal{G}_{[k]}||\mathcal{G}_{[l]}| .$$
$$(18, 19)$$

For $(\Delta\theta^{km}_{ln} + \Delta\phi_{kl}) \ll 1$ we can expand the exponential in equation (19) as $[1 + i(\Delta\theta^{km}_{ln} + \Delta\phi_{kl})]$. In this formulation a number of the results of Miller (1995) become readily apparent. First we expect a shelf in the time lag that is given by a weighted sum of $\Delta\theta^{km}_{ln}/f$. We will see in the next section that $\Delta\theta \sim 2\pi f \bar{\lambda}/c$, where $\bar{\lambda}$ is the mean scattering path length. In order for this to be measurable, we must have $\Delta\theta \gtrsim \Delta\phi \sim 0.1$ radians, or $f \gtrsim 5$ Hz $(\bar{\lambda}/10^8 \text{ cm})^{-1}$ [in agreement with Miller (1995)]. However, in realistic systems, the effects of counting noise introduces an uncertainty of $\sim 0.01 - 0.05$ radians at this frequency, and with GINGA observations of Cygnus X-1 this uncertainty grows *exponentially* with $f$ (*cf.* §5). This is why that in the past the presence or absence of this shelf has not been a useful diagnostic tool. In the above formulation we also see why in Miller's picture we do not alter the functional form of the phase lag. He assumes that $\Delta\theta_{ln}^{km} \ll \Delta\phi_{kl}$ *and* that $\Delta\phi_{kl}$ has the same functional dependence on $f$, no matter what $k$ and $l$ are. Thus, in the small angle approximation, the net phase delay is simply the sum of small constants multiplied by a common function of frequency.

Let us consider simplified versions of equations (18) and (19) by assuming that the redistribution function is so sharply peaked that there is a one-to-one mapping from the input energy band to the output energy band. This assumption is unrealistic; however, it does serve the purpose of ellucidating the basic features of our equations. We shall assume input energy band $[k]$ is mapped into output energy band $[m]$, and that band $[l]$ is mapped into band $[n]$. Equations (18) and (19) then can be rewritten as

$$
\begin{aligned}
|\mathcal{H}_{[m]}|^2 &= \exp\left\{-(f\tau_{km})^2\right\} \; |\mathcal{G}_{[k]}|^2 \; , \\
\mathcal{H}_{[m]}\mathcal{H}_{[n]}^* &= \exp\left\{-\frac{(f\tau_{km})^2 + (f\tau_{ln})^2}{2}\right\} \; \exp\left[i\; \left(\Delta\theta_{ln}^{km} + \Delta\phi_{kl}\right)\right] \; \sqrt{\mathcal{C}_{kl}} \; |\mathcal{G}_{[k]}||\mathcal{G}_{[l]}| \; .
\end{aligned}
\tag{20, 21}
$$

Utilizing the above expressions, we find that the coherence function for the output light curves becomes $\mathcal{C}_{mn} = \mathcal{C}_{kl}$, *i.e.*, it is unchanged from the input coherence function. The exponential degradation of the PSD exactly cancels the exponential degradation of the CPD. In fact one can show that for the general expressions of equations (18) and (19), the coherence function remains identically unity if the input coherence functions are unity. We see that the net effect of a *stationary* Compton cloud is to exponentially degrade the PSD at frequencies $f \gtrsim \tau_{km}^{-1}$, and to introduce a time-lag shelf of amplitude $\Delta\theta_{ln}^{km}/f$. The amplitude of the former effect is determined by the variance of the diffusion time through a cloud. In the next section, we examine likely sources for this variance.

## 3 MEAN AND VARIANCE OF SCATTERING TIMES

The total diffusion time through a Compton cloud is determined by the number of scatters a photon undergoes. This number is in turn determined from the ratio of the output photon energy to that of the input photon energy. Assuming that the range of input photon energies is relatively narrow, the mean number of scatters (and hence the mean diffusion time through the cloud) is fixed by the observed output photon energy. We must then determine the dispersion in this mean diffusion time. In what follows we shall assume a very simple model of Compton scattering, similar to those considered by Sunyaev & Titarchuk (1980), Payne (1980), and others. The goal will be to understand approximately the amount of variance in the diffusion times that one can expect from "typical" Compton cloud models.

Following Sunyaev & Titarchuk (1980), we define the Compton time parameter

$$
u = \kappa_{es}\rho \; ct \; , \tag{22}
$$

which is essentially the mean number of scatters a photon undergoes in a time $t$. The mean free path of the photon is given by $\bar{\lambda} = (\kappa_{es}\rho)^{-1}$. For a number of different source distributions the probability of a photon escaping the cloud goes as $P(u) \approx \bar{N}^{-1} \exp(-u/\bar{N})$, with $\bar{N} \propto \tau_{es}^2$, where $\tau_{es}$ is the electron scattering optical depth (*cf.* Sunyaev & Titarchuk 1980). The exponential decrease in this probability is balanced by the exponential increase (with number of scatters) of the photon energy, resulting in a characteristic power law spectra with photon number index $\sim -2$. We shall take the increase in photon energy to go as

$$\nu = \nu_0 \exp\left(3\frac{kT}{mc^2}u\right) \; , \qquad (23)$$

where $\nu_0$ is the input photon frequency, $kT$ is the electron temperature of the cloud, and $mc^2$ is the rest mass energy of the electron (*cf.* Pozdnyakov et al. 1983). The number of scatters therefore goes as

$$u = \frac{1}{3}\frac{mc^2}{kT}\ln(\nu/\nu_0) \; . \qquad (24)$$

If we assume that the input photons have energy of 150 eV and that the output photons are upscattered in a 50 keV cloud to $\sim 5$ keV, then the photons undergo approximately 10 scatters. The mean diffusion time is given by $u\bar{\lambda}/c$, and therefore scales as $\ln(\nu/\nu_0)$. If one assumes that all observed photons originate at the same energy, then the difference in diffusion times required to obtain energies $E_1$ and $E_2$ is proportional to $\ln(E_1/E_2)$. The shelf in the time lag is given by this difference, and hence has this same scaling (Miller 1995).

There are a number of possible sources of variance in the diffusion time. Assuming a relatively narrow spectrum of input photons, there are four likely sources of variance: finite width of the output energy bands leading to multiple diffusion times being represented in a single band; geometrical effects leading to different photon arrival times; dispersion in the number of scatters required to attain a given output energy; variation in the path length travelled by a photon inbetween scatters. Of all of these effects, variation in the path length will be seen to be the most important.

The first two effects, finite energy bands and geometrical effects, are the least important. A single scatter on average produces a frequency change of

$$\left\langle \frac{\nu'}{\nu} \right\rangle = 1 + 4\frac{kT}{mc^2} \; , \qquad (25)$$

where $\nu$ and $\nu'$ are the photon frequencies before and after scattering, respectively (*cf.* Pozdnyakov et al. 1983). If we take $kT \sim 50$ keV, then each scatter on average produces an energy change of $\sim 40\%$. For typical X-ray bands, the difference between the extremes of the energy band and the average photon energy in a band is $\lesssim \pm 20\%$, *i.e.* $\sim \pm 0.5$ scatters. Thus, the variance in the diffusion time is $\sim 0.5 \; \bar{\lambda}/c$. As an estimate of the geometrical effects, let us consider a point source embedded within a uniformly bright spherical cloud of radius $R$. Approximately 68% of the emergent flux comes from within an impact parameter of $\lesssim 0.8 \; R$, which leads to a time delay (between impact parameters 0 and 0.8 $R$) of $\sim 0.4 \; R/c$. Since $R \sim \tau_{es}\bar{\lambda}$, and for most models $\tau_{es} \sim 1-3$, the dispersion is $\sim 0.4 - 1.2 \; \bar{\lambda}/c$.

We shall now make a very simple estimate of the dispersion in the number of scatters required to reach a given energy. In the same manner one can calculate the mean change in frequency per scatter, one can show that

$$\left\langle \left(\frac{\nu'}{\nu}\right)^2 \right\rangle \approx 1 + 10\frac{kT}{mc^2} \; , \qquad (26)$$

(in the non-relativistic limit). The variance in the frequency shift is thus of $\mathcal{O}(2kT/mc^2)$. As a simple estimate, the ratio of the input photon energy, $E'$, to the output photon energy, $E$ is approximately given by

$$\ln\left(\frac{E}{E'}\right) \sim N\left(\frac{4kT}{mc^2}\right) \pm \sqrt{N}\left(\frac{2kT}{mc^2}\right) , \qquad (27)$$

where $N$ is the number of scatters that the photon undergoes. By the central limit theorem, the above distribution is approximately Gaussian. Setting $x \equiv E/E'$, the joint probability distribution of $x$ and $N$ is then approximately

$$P(x, N) \propto \exp\left(-\frac{[\ N\ -\ (mc^2/4kT)\ln x\ ]^2}{N/2}\right) . \qquad (28)$$

For $N$ sufficiently large, the above distribution is sharply peaked and is approximately Gaussian with standard deviation $\sigma \approx \sqrt{N}/2$. The dispersion in the time delay is then $\sim \sqrt{N}\,\bar{\lambda}/2c$, which for $N \sim 10$ is $\sim 1.6\,\bar{\lambda}/c$. In general, this contribution is larger than the finite energy band and geometric effects discussed above, and becomes increasingly important as one considers higher photon energies (*i.e.*, larger $N$).

The dominant source of variance in the diffusion time is variation in the path length travelled between scatters. The mean free path is $\bar{\lambda}$, however, the standard deviation of the path length is also $\bar{\lambda}$. After $N$ scatters, the total path length travelled will be distributed approximately as a Gaussian with mean $N\,\bar{\lambda}$ and standard deviation $\sigma = \sqrt{N}\,\bar{\lambda}$. The variance in the time delay is then given by $\sqrt{N}\,\bar{\lambda}/c$, or $\sim 3.2\,\bar{\lambda}/c$ for $N = 10$. Adding all the sources of variance together in quadrature (for $N = 10$, $\tau_{es} = 3$), we obtain a total variance of $\sim 3.8\,\bar{\lambda}/c$. That is, the variance is dominated almost entirely by the standard deviation of the path length and the variation in the energy. Both are approximately $\sqrt{N}$ effects and are approximately Gaussianly distributed. In the next section we will continue to use the assumption of a Gaussian distribution in the diffusion time.

## 4 STATIC CLOUD MODELS

Following Miller (1995), we shall investigate the effects of a static Compton cloud on an assumed form of the input variability. Specifically, we shall assume that the input spectrum is a blackbody with a PSD $\propto f^{-1}$. Furthermore, the amplitude of the PSD is assumed to be independent of input photon energy, and only proportional to the square of the number of photons in a given input energy band. In turn, we choose the input energy bands in such a way that there is an equal number of photons per band. (In practice we have found that relatively few input energy bands – here we use twenty one – is sufficient to guarantee that the results do not change by increasing the number of bands.) In what follows, we use three different input blackbody temperatures: 0.15 keV, 0.6 keV, and 1.5 keV. These cover a wide range of the "typical" input spectra that are considered when Compton scattering models are fit to X-ray and gamma-ray data (*cf.* Titarchuk 1994).

We also must choose a Fourier phase for the input photons. Here we choose a simpler functional form than that of Miller (1995). Cygnus X-1 shows a phase difference between the hard and soft photons that is nearly a constant 0.1 radians. This would imply a time lag proportional to $f^{-1}$ (in reality, $f^{-2/3}$ is a better fit; Miyamoto et al. 1988). We therefore adopt a frequency independent phase difference of 0.1 radians between the seventh energy band and the fifteenth energy band.

The lower band is given a fixed phase of $\pi/2$, and then the phase is extrapolated to other bands by assuming that it is proportional to the logarithm of the mean band energy. [This is similar to the functional dependence chosen by Miller (1995).]

For our Compton cloud model we use the simple analytic model of the previous section, with $kT = 50$ keV and $\tau_{es} = 3$. We chose these parameters so that we could directly compare our resuts to those of Miller (1995). The redistribution matrices, the mean photon diffusion time, and the variance of the mean diffusion time were all via the equations of §3. Utilizing equations (18) and (19), we then generated output PSD and time lags for the two output bands $1.0 - 3.2$ keV and $12.1 - 60$ keV. These bands represent the lowest and highest energy bands that one can realistically expect to use with the XTE satellite. We chose three values for $\bar{\lambda}$: $1 \times 10^6$ cm, $1 \times 10^7$ cm, and $1 \times 10^8$ cm. Output time lags for these nine runs are shown in Figure 1. Note that for this figure we have taken the intrinsic coherence to be unity between all input energy bands. That is, $\mathcal{C}_{[k\,l]} = 1$ for all $k$ and $l$. The associated PSDs, assuming an $f^{-1}$ input, are shown in Figure 2.

For the most part, our results are in qualitative agreement with those of Miller (1995). That is, the functional form of the time lags (for us, $\propto f^{-1}$) is preserved by the scattering. In addition, a "shelf" is introduced into the time lag at high frequencies (short periods). The level of this shelf increases as the scattering path length is increased. (The sudden drop-off in this shelf is due to the fact that, by definition, the phase lag is constrained to the interval $-\pi \to \pi$. Once the phase due to the shelf becomes greater than $\pi$, it is mapped to a negative angle.) Not shown in Miller (1995) are the effects of scattering on the PSD (though see Payne 1980, Kylafis & Klimis 1987, Brainerd et al. 1987, Bussard et al. 1988, Miller & Lamb 1992). The exponential rollover is readily apparent for $\bar{\lambda} \sim 1 \times 10^8$ cm. In addition, Miller (1995) did not show the noise limits, which we show as dashed lines in Figures 1 and 2. (For the phase lags we chose a noise limit consistent with previous Ginga observations of Cygnus X-1; *cf.* Vaughan 1991. For the PSDs we chose a noise limit consistent with that expected for XTE observations.) Note that the time lag noise limit sharply rises at high frequency. This is due to the exponential loss of coherence at high frequency (*cf.* §5 and Vaughan 1991), which was seen in Ginga data of Cygnus X-1. The net effect of this coherence loss was to make it very difficult to discern the presence or absence of a time lag shelf. The observed PSD is typically somewhat higher above the effective noise level than the observed time lag. (Essentially this is because there exists an "optimal filter" for deconvolving the noise from the PSD and the *amplitude* of the CPD, but there is no such filter for deconvolving the CPD phase; *cf.* Davenport & Root 1987, Vaughan 1991.) It is therefore somewhat easier to search for an exponential rollover in the PSD than it is to search for a time lag shelf. The lack of an observed exponential rollover limits $\bar{\lambda} \lesssim 1 \times 10^8$ cm (for this particular Compton cloud model).

The one notable difference between our results and those of Miller (1995) is for the output time lags of low energy input photons. Specifically, we see that for an input blackbody of $kT = 0.15$ keV the intrinsic constant phase lag of 0.1 radians is nearly entirely wiped out. All that remains for the most part is the time lag shelf due to the diffusion time through the cloud. The reason for this is simple. For any two input channels, $A$ and $B$, being scattered into output channels, 1 and 2, we have to consider the scattering processes $A \to 1$, $B \to 2$ *and* $A \to 2$, $B \to 1$. These processes enter into equation (19) with equal and opposite phases. Therefore, if $R_{[A1]}R_{[B2]} \approx R_{[A2]}R_{[B1]}$, then the intrinsic phases cancel. This condition is approximately true when the output photon energy is much greater than the input photon energy. On the other hand, if these energies are comparable, then the condition does not hold. (If $A,B$ and $1,2$ are each in order of increasing energy, $R_{[B1]}$ tends to be $\sim 0$, since it is highly unlikely that a hot Compton cloud will downscatter photons.) Miller (1995) considered output channels of mean energy 2.5 and 5.0 keV with a 2.5 keV blackbody input; therefore, he did not detect any phase cancellation. This cancellation of the intrinsic phase at high energies is a potentially useful diagnostic tool to determine whether or not the observed variability is consistent with that of an intrinsic source being passed through a Compton cloud.

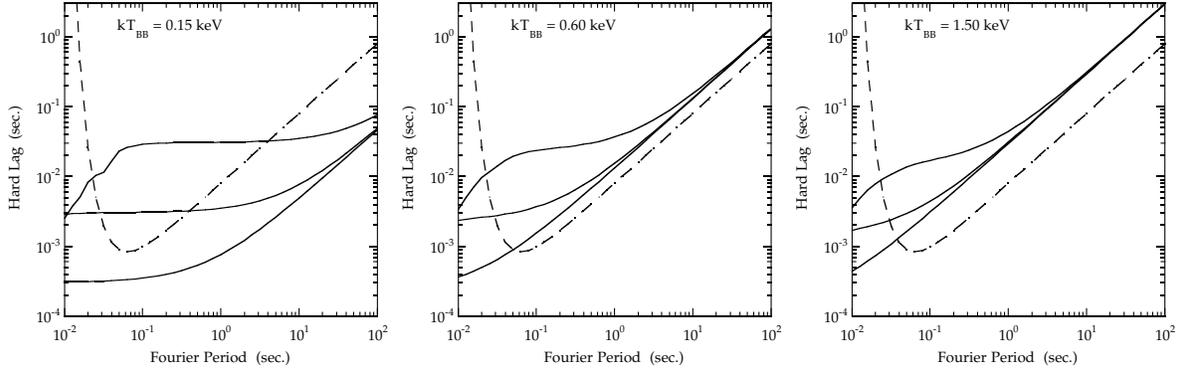

**Figure 1.** Left: Effects of Compton scattering on intrinsic time lags between hard and soft photons (taken to be given by a 0.1 radians phase delay between the bottom and top third input energy bands for input blackbodies with, from left to right, $kT_{BB} = 0.15$ keV, 0.60 keV, 1.5 keV). Dashed lines show noise estimates appropriate for Ginga data (*cf.* Vaughan 1991). In all cases, the soft output energy band was taken to be $1 - 3.2$ keV and the hard band was taken to be $12.1 - 60$ keV. The mean free scattering path lengths were taken to be $\bar{\lambda} = 10^6$ cm (bottom solid lines), $10^7$ cm (middle solid lines), and $10^8$ cm (top solid lines).

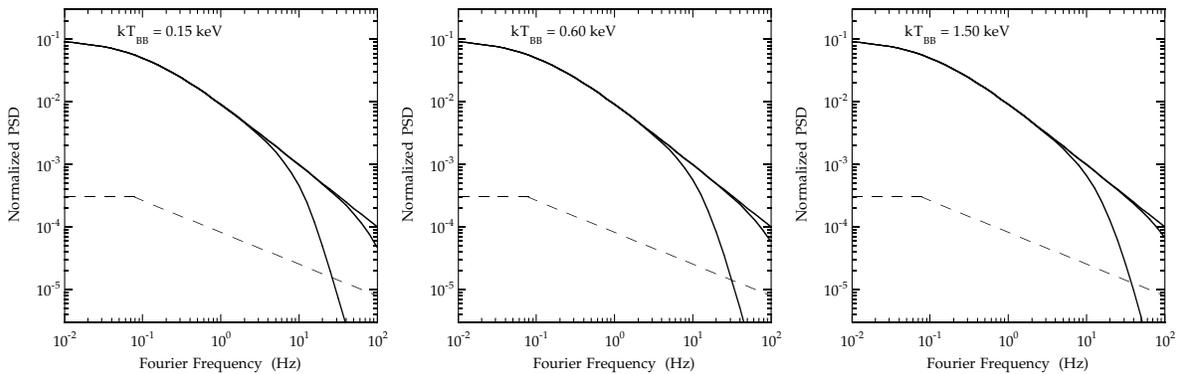

**Figure 2.** Effects of Compton scattering on an intrinsic $1/f$ PSD (with flattening for $f < 0.1$ Hz). The models are identical to Figure 1. The dashed line represents the expected XTE noise limit for the $12.1 - 60$ keV band. The mean free scattering path lengths again were taken to be $\bar{\lambda} = 10^6$ cm (top solid lines), $10^7$ cm (middle solid lines), and $10^8$ cm (bottom solid lines).

## 5 NON-STATIONARY CLOUD MODELS

In the previous section, we considered the case of a stationary cloud, so that equations (18) and (19) were the valid expressions for deriving the output variability from the input variability. Here we shall relax the stationary assumption somewhat. We shall still assume that for an *individual* time series [denoted with a superscript $i$ in equations (1) − (10)] the probability distribution, equation (5), is well-defined. However, we shall now assume that the mean diffusion time and the variance of this diffusion time varies from sample to sample. That is, we shall assume that these quantities are themselves statistically distributed. For example, we might imagine that the temperature of the Compton cloud fluctuates about some mean value. These fluctuations will in turn lead to fluctuations in the mean diffusion time, the standard deviation of this time, and in the redistribution function. Or we might imagine that there are fluctuations in the physical size of the cloud that leads to fluctuations in the mean free scattering path, $\bar{\lambda}$. Such fluctuations would cause the mean diffusion time and its standard deviation to vary.

It is important to note that we are considering these fluctuations to be occuring on time scales longer than the individual time series that are used to calculate the mean output PSD and CPD. For typical observations, the lengths of these time series are $10 - 100$ seconds. Fluctuations that occur on time scales much less than $\Delta$, the bin width of the time series, will tend to average out and leave us with a scattering cloud that appears stationary. Fluctuations on time scales greater than $\Delta$ but less than the length of the individual time series invalidates the approximations of equation (9), and will lead to a more complicated formalism than that presented here. For the time scales with which we are concerned, equations (18) and (19) hold for *individual* time series, with $\mathcal{G}$, $\tau_{km}$, $\Delta\theta_{lm}^{km}$, $R_{[km]}$, etc., now considered to be random variables. We find the mean output PSD and CPD by averaging over the probability distributions of these variables.

As a simple example, let us imagine that the mean free scattering path is given by $\bar{\lambda} = a\langle\bar{\lambda}\rangle$, where $\langle\bar{\lambda}\rangle$ is the average about which the mean free path fluctuates, and $a$ is a random variable distributed about a mean of 1. We then see that $\Delta\theta_{lm}^{km} = a\langle\Delta\theta_{lm}^{km}\rangle$, $\tau_{km} = a\langle\tau_{km}\rangle$, etc. The redistribution function is not effected by variations in the mean free path. Now let us assume that $a$ is Gaussianly distributed with standard deviation $\sigma_a$. Note that we must have $\sigma_a \ll 1$ (negative values of $\bar{\lambda}$ are unphysical). We can now average equations (18) and (19) over this distribution. The general form of the equations remain unchanged; however, this averaging leads to the following substitutions. Terms of the form $\exp[-(f\tau)^2]$ are replaced by

$$\frac{1}{\sqrt{1+2(\sigma_a f\tau)^2}} \exp\left[-\frac{(f\tau)^2}{1+2(\sigma_a f\tau)^2}\right] \quad . \tag{29}$$

Terms of the form $\exp(i\,\Delta\theta)$ are replaced by

$$\exp\left(-\frac{(\sigma_a \Delta\theta)^2}{2[1+2(\sigma_a f\tau)^2]}\right) \exp\left[\frac{i\,\Delta\theta}{1+2(\sigma_a f\tau)^2}\right] \quad . \tag{30}$$

The $\cos(\Delta\theta)$ terms are replaced in exactly the same manner as the $\exp(i\,\Delta\theta)$ terms.

There are several interesting effects to notice here. First, at high frequencies the exponential decay of the PSD and CPD is replaced by a more gradual power law decay. Second, the time lag shelf is less pronounced, being reduced by a factor $[1+2(\sigma_a f\tau)]$. Third, the CPD and to a lesser extent the PSD, are reduced due terms of the form of equation (30). These terms lead to a net loss of coherence that at first is seen to decay exponentially, but then levels out to a constant reduction factor.

Examples of this coherence loss are shown in Figure 3. We take the same Compton cloud models as in Figures 1 − 2, and we input a 0.6 keV blackbody in an identical manner as that discussed in §4. The average mean free scattering path is taken to be $\langle \bar{\lambda} \rangle = 10^8$ cm, and the normalized standard deviation is taken to be $\sigma_a = 0.3$, 0.5. Figure 5 shows that the coherence begins to deviate from unity above 2 Hz. This is qualitatively similar to what has been observed to occur in the high frequency coherence of Cygnus X-1 (Vaughan 1991); however, in that particular case the coherence was seen to be degraded even more strongly. The coherence loss seen in the GINGA data may imply unphysically large cloud variations; however, a large fraction of that coherence loss may be due to deadtime effects in the GINGA detectors (Vaughan 1995, unpublished). XTE, with its higher frequency coverage and lower deadtime should be able to measure the coherence function out to ∼ 100 Hz, and therefore place interesting limits on Compton cloud variability in Cygnus X-1.

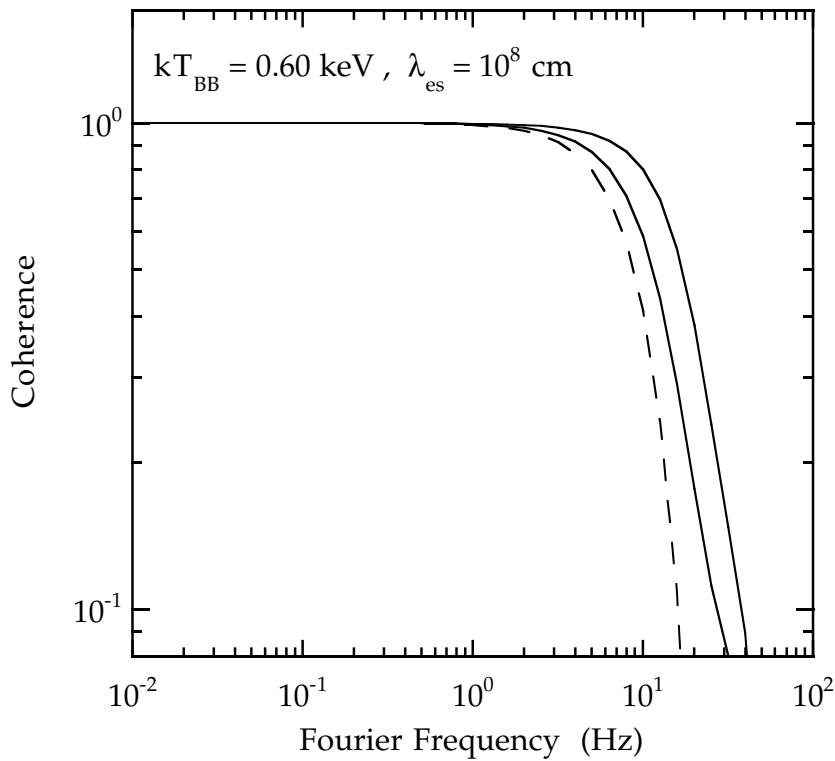

**Figure 3.** Coherence function for the comparison of 1 − 3.2 keV ouptut photons with 12.1 − 60 keV output photons. The Compton cloud model was the same as that used in Figures 1 & 2, with an input blackbody spectrum of $kT = 0.6$ keV. The average mean free scattering path was taken to be $\langle \bar{\lambda} \rangle = 10^8$ cm, and the normalized variance of the path length was taken to be $\sigma_a = 0.3$ (top line) and $\sigma_a = 0.5$ (bottom line). The dashed line shows the typical coherence loss seen in Ginga data of Cygnus X-1 (Vaughan 1991).

Instead of assuming that $\bar{\lambda}$ fluctuated, we could have also assumed that the cloud temperature fluctuated. This would lead to qualitatively similar results as those discussed above; however, we also would have to evaluate the correlated fluctuations in the redistribution matrices $R_{[kl]}$. These calculations are not difficult in principle, and would be most easily accomplished via a numerical Compton cloud model. The large drop in coherence observed in Cygnus X-1 would then indicate commensurately large variations in the temperature of the Compton cloud in this system.

## 6 SUMMARY

In this paper we have examined the claim of Miller (1995) that scattering in an optically thick Compton cloud does not grossly change the phase lags between hard and soft photons intrinsic to the soft input source. Though this was true when the energy of the input photons was comparable to that of the output photons, the intrinsic phases cancelled whenever the input energy was much less than the output energy. In both cases there was a time lag "shelf" whose amplitude was dependent upon differences in the mean diffusion time through the cloud. The amplitude scaled as $\ln(E_1/E_2)$, where $E_1$ and $E_2$ were the energies of the output bands that were to be compared [as also pointed out by Miller (1995)]. Unfortunately, the region of this shelf has been obscured by Poisson noise for previous observations with satellites such as GINGA.

In addition to the potential phase cancellation, we noted that scattering will lead to a degradation of the PSD. The degree to which the variability is reduced was seen to depend upon the variance in the diffusion time, which in turn was seen to scale as the square root of the number of mean scatters. This rollover in the PSD is potentially more observable than the time lag shelf since it is somewhat easier to remove noise effects from PSD estimates than it is to remove noise from phase measurements.

We also defined the coherence function between two measured energy bands. Essentially it is the normalized amplitude of the mean cross power spectral density, and it is a measure of the correlation between the variability – at a given frequency – of two energy bands (Vaughan 1991). If the intrinsic coherence between all input energy bands is unity, then for a stationary Compton cloud the coherence between the output energy bands will also be unity. We saw, however, that a fluctuating Compton cloud could lead to a loss of coherence at high frequency. In order for the loss of coherence to be significant the variation in the cloud parameters must be large. Coherence loss at high frequency has been detected in Cygnus X-1, but for this particular case the coherence loss implies unphysically large cloud fluctuations. However, much of this coherence loss may be attributable to deadtime effects in the GINGA detectors. XTE, with its ability to probe high frequencies with low deadtime should be able to accurately measure the coherence functionin Cygnus X-1 out to $\sim 100$ Hz.

Miller (1995) argued that if variations in the Compton cloud are small, then they play a negligible roll in creating Fourier frequency dependent phase lags. On the other hand, the coherence function rollover seen in Cygnus X-1 – if real – may be indicating that cloud fluctuations are indeed large. This opens up the possibilty that the observed phase lags are produced by the cloud itself. If a Compton cloud itself produces the phase lags there then is also no worry of the cloud wiping out the intrinsic phase lags of the soft input source. If true, observations of phase lags and coherence in sources such as Cygnus X-1 are telling us more about the physical conditions within Compton clouds rather than about the structure of the soft input source.

## ACKNOWLEDGEMENTS

We would like to thank Cole Miller for providing us with a preprint of his work before publication, and for useful correspondence. We would also like to thank Peter Michelson and Chris Thompson for useful conversations. M.A.N. was supported by NASA Grant NAGW-4484 and B.A.V. was partly supported by the Netherlands Organization for Scientific Research (NWO) under grant PGS 78-277.